\documentclass[twocolumn ,showpacs  ,secnumarabic, nobibnotes,  nofootinbib,
aps, prd]{revtex4}%
\usepackage{amsmath}
\usepackage{amssymb}
\usepackage{bm}
\usepackage{graphicx}
\usepackage{amsfonts}%
\providecommand{\U}[1]{\protect\rule{.1in}{.1in}}

\begin{document}
\title{Kinematics of deep inelastic scattering in leading order of 
the covariant approach }
\author{Petr Zavada}
\email{zavada@fzu.cz}
\affiliation{Institute of Physics AS CR, Na Slovance 2, CZ-182 21 Prague 8, Czech Republic}

\begin{abstract}
We study the kinematics of deep inelastic scattering corresponding to the
rotationally symmetric distribution of quark momenta in the nucleon rest
frame. It is shown that rotational symmetry together with Lorentz invariance
can in leading order impose constraints on the quark intrinsic momenta.
Obtained constraints are discussed and compared with the available
experimental data.

\end{abstract}

\maketitle

\section{Introduction}

\label{sec1}The motion of quarks inside the nucleons plays an important role
in some effects which are at present intensively investigated both
experimentally and theoretically. Actual goal of this effort is to obtain a
more consistent 3-D picture of the quark-gluon structure of nucleons. For
example the quark\ transversal momentum creates the asymmetries in particle
production in polarized (SIDIS) or in unpolarized (Cahn effect) experiments on
deep inelastic scattering (DIS). Relevant tool for the study of these effects
is the set of the transverse momentum dependent distributions (TMDs).
Apparently, a better understanding of the quark intrinsic motion is also a
necessary condition to clarify the role of \ quark orbital angular momenta in
generating nucleon spin.

We have paid attention to these topics in our recent studies, see
\cite{Efremov:2010mt,Zavada:2009sk,Efremov:2010cy,Efremov:2009ze,Zavada:2007ww}
and citations therein. In particular we have shown that the requirements of
Lorentz invariance (LI) and the nucleon rotational symmetry in its rest frame
(RS), if applied in the framework of the 3-D covariant quark-parton model
(QPM), generate a set of relations between parton distribution functions.
Recently we obtained within this approach relations between usual parton
distribution functions and the TMDs. The Wanzura-Wilczek approximate relation
(WW) and some other known relations between the $g_{1}$ and $g_{2}$ {structure
functions were similarly obtained in the same model before
\cite{Zavada:2002uz}. Let us remark that the WW relation has been obtained
independently also in another approaches \cite{Jackson:1989ph,
D'Alesio:2009kv} in which the LI represents a basic input.}

The aim of the present report is to consistently apply the assumption LI\&RS
to the kinematics of DIS and to obtain the constraints on related kinematical
variables. That is a complementary task to the study of above mentioned
relations between distribution functions, which depend on these variables. So,
the report can be considered as an addendum to our former papers related to
the covariant QPM
\cite{Efremov:2010mt,Zavada:2009sk,Efremov:2010cy,Efremov:2009ze,Zavada:2007ww,Efremov:2004tz,Zavada:2002uz,Zavada:2001bq,Zavada:1996kp}%
. \qquad

Since the present discussion is motivated and based on our earlier study of a
covariant version of QPM, obtained results correspond only to the leading
order of a more consistent QCD treatment. In this sense it would be
interesting to compare our results with the leading order of a real QCD
approach, like e.g. the recent study of perturbative QCD evolution of TMDs
{\cite{arXiv:1101.5057, Aybat:2011ta}}. However such task would go beyond the
scope of this short report. Anyway, in general a comparison between the
experimental data and the leading order predictions can be important and instructive.

\section{Kinematic variables}

\subsection{The Bjorken variable and light-cone coordinates}

\begin{figure}[ptb]
\includegraphics[width=5cm]{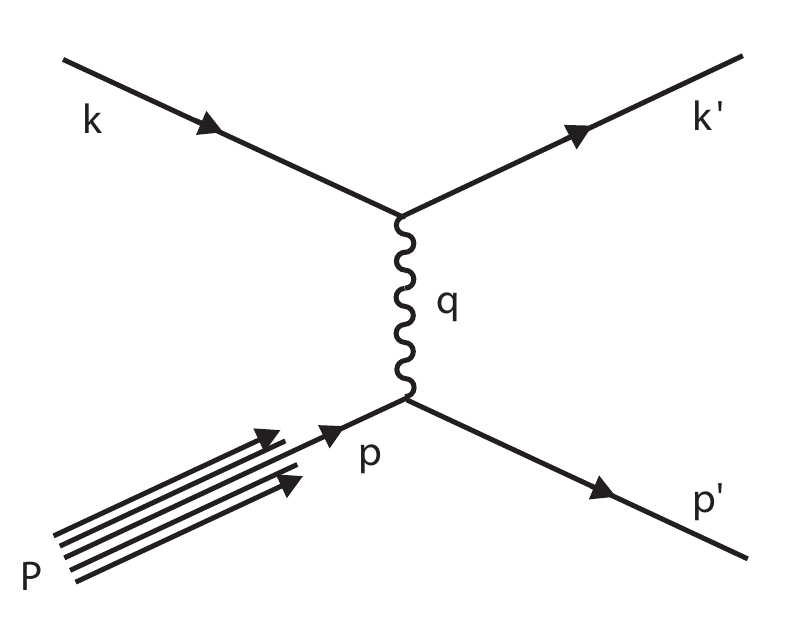}\caption{Diagram describing DIS as a
one photon exchange between the charged lepton and quark. Lepton and quark
momenta are denoted by $k,p$ ($k^{\prime},p^{\prime}$) in initial (final)
state, $P$ is initial nucleon momentum.}%
\label{fe1}%
\end{figure}

First, let us shortly remind the properties of the Bjorken variable%
\begin{equation}
x_{B}=\frac{Q^{2}}{2Pq}, \label{k1}%
\end{equation}
which plays a crucial role in phenomenology of lepton -- nucleon scattering.
Regardless of mechanism of the process, this invariant parameter satisfies%
\begin{equation}
0\leq x_{B}\leq1. \label{k2}%
\end{equation}
This is a very well-known textbook result. A possible proof is suggested also
in {\cite{arXiv:1106.5607}.} Now let us consider a QPM approach, where the
process of lepton -- nucleon scattering is initiated by the lepton interaction
with a quark (see Fig. \ref{fe1}), which obeys
\begin{equation}
p^{\prime}=p+q,\qquad p^{\prime2}=p^{2}+2pq-Q^{2};\qquad Q^{2}=-q^{2}.
\label{k3}%
\end{equation}
The second equality implies%
\begin{equation}
Q^{2}=2pq-\delta m^{2};\qquad\delta m^{2}=p^{\prime2}-p^{2}, \label{k4}%
\end{equation}
which with the use of relation (\ref{k1}) gives%
\begin{equation}
\frac{pq}{Pq}=x_{B}\left(  1+\frac{\delta m^{2}}{Q^{2}}\right)  . \label{k5}%
\end{equation}
The basic input for the construction of QPM is the assumption
\begin{equation}
Q^{2}\gg\delta m^{2}, \label{k6}%
\end{equation}
which allows us to identify%
\begin{equation}
x_{B}=\frac{Q^{2}}{2Pq}=\frac{pq}{Pq} \label{k7}%
\end{equation}
and to directly relate the quark momentum to the parameters of scattered
lepton. Moreover, if one assumes%
\begin{equation}
Q^{2}\gg4M^{2}x_{B}^{2}, \label{k8}%
\end{equation}
where $M$ is the nucleon mass, then one can identify%
\begin{equation}
x_{B}=x\equiv\frac{p_{0}-p_{1}}{P_{0}-P_{1}} \label{k9}%
\end{equation}
in any reference frame in which the first axis orientation is defined by the
vector $\mathbf{q}$. This relation can be proved as follows. Let us consider
Eq. (\ref{k7}) in the same frame:%
\begin{equation}
x_{B}=\frac{p_{0}q^{0}-p_{1}\left\vert \mathbf{q}\right\vert }{P_{0}%
q^{0}-P_{1}\left\vert \mathbf{q}\right\vert }. \label{ax6}%
\end{equation}
In the nucleon rest frame we denote the photon momentum components by the
subscript $R$ and using the usual symbol $\nu=q_{R}^{0}$ we have%
\begin{equation}
\left\vert \mathbf{q}_{R}\right\vert ^{2}=\nu^{2}+Q^{2}, \label{ax7}%
\end{equation}
which with the use of Eq. (\ref{k1}) gives%
\begin{equation}
\frac{\left\vert \mathbf{q}_{R}\right\vert ^{2}}{\nu^{2}}=1+\frac{4M^{2}%
x_{B}^{2}}{Q^{2}}. \label{ax7a}%
\end{equation}
It means that for $Q^{2}\gg4M^{2}x_{B}^{2}$ we obtain%
\begin{equation}
\frac{\left\vert \mathbf{q}_{R}\right\vert }{\nu}=1+O\left(  \frac{4M^{2}%
x_{B}^{2}}{Q^{2}}\right)  . \label{ax7b}%
\end{equation}
(Since for $Q^{2}\rightarrow\infty$ we have also $\left\vert \mathbf{q}%
_{R}\right\vert ,\nu\rightarrow\infty,$ so the ratio $\left\vert
\mathbf{q}_{R}\right\vert /\nu\rightarrow1$ does not contradict Eq.
(\ref{ax7}).) In a reference frame connected with the rest frame by the
Lorentz boost in the direction\ opposite to $\mathbf{q}_{R}$ we have the
corresponding ratio%
\begin{equation}
\frac{q^{1}}{q^{0}}=\frac{\left\vert \mathbf{q}_{R}\right\vert +\beta\nu}%
{\nu+\beta\left\vert \mathbf{q}_{R}\right\vert }. \label{ax7c}%
\end{equation}
Now one can easily check that Eq. (\ref{k7}) with the use of this ratio and
Eq. (\ref{ax7b}) imply
\begin{equation}
x_{B}=\frac{p_{0}q^{0}-p_{1}q^{1}}{P_{0}q^{0}-P_{1}q^{1}}=\frac{p_{0}-p_{1}%
}{P_{0}-P_{1}}\left(  1+O\left(  \frac{4M^{2}x_{B}^{2}}{Q^{2}}\right)
\right)  . \label{ax8}%
\end{equation}
In this way we have proved that replacement of Bjorken variable by the
invariant light-cone ratio in Eq. (\ref{k9}) is valid provided the inequality
(\ref{k8}) is satisfied.

The relation (\ref{k9}) expressed in the nucleon rest frame reads%
\begin{equation}
x=\frac{p_{0}-p_{1}}{M}, \label{k10}%
\end{equation}
which after inserting into (\ref{k2}) gives%
\begin{equation}
0\leq\frac{p_{0}-p_{1}}{M}\leq1. \label{k11}%
\end{equation}

However the most important reason why we require large $Q^{2}$ is in physics.
If we accept scenario when a probing photon interact with a quark, we need
sufficiently large momentum transfer $Q^{2}$ at which the quarks can be
considered as effectively free due to asymptotic freedom. At small $Q^{2}$ the
picture of quarks (with their momenta and other quantum numbers) inside the
nucleon disappear.

\subsection{Rotational symmetry}

The RS means that the probability distribution of the quark momenta
$\mathbf{p}=(p_{1},p_{2},p_{3})$\ in the nucleon rest frame depends, apart
from $Q^{2}$, on $\left\vert \mathbf{p}\right\vert $. It follows that also
$-\mathbf{p}$ is allowed, so together with the inequality\ (\ref{k11}) we have%
\begin{equation}
0\leq\frac{p_{0}+p_{1}}{M}\leq1. \label{k12}%
\end{equation}
The combinations of (\ref{k11}),(\ref{k12}) imply%
\begin{equation}
0\leq\left\vert p_{1}\right\vert \leq p_{0}\leq M,\qquad\left\vert
p_{1}\right\vert \leq\frac{M}{2}. \label{k14}%
\end{equation}
And if we again refer to RS, then further inequalities are obtained:%
\begin{equation}
0\leq\left\vert \mathbf{p}\right\vert \leq p_{0}\leq M,\quad\left\vert
\mathbf{p}\right\vert \leq\frac{M}{2},\quad0\leq p_{T}\leq p_{0}\leq M
\label{k15}%
\end{equation}
and%
\begin{equation}
p_{T}\leq\frac{M}{2}, \label{k15a}%
\end{equation}
where%
\[
\left\vert \mathbf{p}\right\vert =\sqrt{p_{1}^{2}+p_{2}^{2}+p_{3}^{2}},\qquad
p_{T}=\sqrt{p_{2}^{2}+p_{3}^{2}}.
\]
Obviously inequality (\ref{k15a}) is satisfied also in any reference frame
boosted in the directions $\pm\mathbf{q}$. Further, the above inequalities are
apparently valid also for average values $\left\langle p_{0}\right\rangle
,\left\langle p_{1}\right\rangle ,\left\langle \left\vert \mathbf{p}%
\right\vert \right\rangle $ and $\left\langle p_{T}\right\rangle $. In
addition, if one assumes that $p_{T}-$ distribution is a decreasing function,
then necessarily%
\begin{equation}
\left\langle p_{T}\right\rangle \leq\frac{M}{4}. \label{k16}%
\end{equation}
\qquad\qquad The above relations are valid for sufficiently high $Q^{2}$
suggested by Eqs. (\ref{k6}) and \ (\ref{k8}). Let us note that the
on-mass-shell assumption has not been applied for obtaining these relations.

These inequalities can be compared with relations obtained in
\cite{Zavada:1996kp}, where the additional on-mass-shell condition
$m^{2}=p^{2}=p_{0}^{2}-\mathbf{p}^{2}$ had been applied. Corresponding
relations are more strict:%
\begin{equation}
\frac{m^{2}}{M^{2}}\leq x\leq1,\quad p_{0}\leq\frac{M^{2}+m^{2}}{2M}%
,\quad\left\vert \mathbf{p}\right\vert \leq\frac{M^{2}-m^{2}}{2M} \label{k13}%
\end{equation}
and%
\begin{equation}
p_{T}^{2}\leq M^{2}\left(  x-\frac{m^{2}}{M^{2}}\right)  \left(  1-x\right)  .
\label{k17}%
\end{equation}
However, it is clear that in general the on-mass-shell assumption is not
realistic. In the next we will assume only the off-mass-shell approach.

\section{Discussion}

First let us summarize more accurately what we have done in the previous
section. We assumed:

\textit{a) Lorentz invariance\newline}It means that the theoretical
description in terms of the standard kinematical variables (see Fig.
\ref{fe1})%
\[
q,\ x_{B},\ x,\ p=(p_{0},p_{1},p_{2},p_{3}),\ P=(P_{0},P_{1},P_{2},P_{3})
\]
\ can be boosted also to the nucleon rest frame.

\textit{b) Inequality }$0\leq x\leq1$\textit{\newline}It means that the
light-cone ratio $x$ satisfies the same bound (\ref{k2}) as the Bjorken
variable $x_{B}.$

\textit{c) Rotational symmetry\newline}The kinematical region $\mathcal{R}%
^{3}$ of the quark intrinsic momenta $\mathbf{p}=(p_{1},p_{2},p_{3})$ in the
nucleon rest frame has rotational symmetry (i.e. $\mathbf{p\in}\mathcal{R}%
^{3}\Rightarrow\mathbf{p}^{\prime}=\mathbf{Rp}\in\mathcal{R}^{3}$, where
$\mathbf{R}$ is any rotation in $\mathcal{R}^{3}$). For example, in terms of
the covariant QPM it means that probabilistic distribution of the quark
momenta is controlled by some function $G\left(  pP/M,Q^{2}\right)  $.

We proved\ these assumptions imply bounds (\ref{k11})--(\ref{k15a}). Now we
will shortly comment on the obtained results:

\textit{i) }The ratio $x$ of light-cone variables (\ref{k9}) has a simple
interpretation in a frame, where the proton momentum is large - $x$ is the
fraction of this momentum carried by the quark. However an interpretation of
the same variable in the nucleon rest frame is more complicated. In this frame
the quark transversal momentum cannot be neglected and $x$ depends on the
both, longitudinal and transversal quark momenta components. In the limit of
massless quarks the connection between the variable $x$ in (\ref{k10}) and the
quark momenta components is given by the relations:
\begin{align}
x  &  =\frac{p_{0}-p_{1}}{M};\qquad\qquad\qquad p_{0}=\sqrt{p_{1}^{2}%
+p_{T}^{2}},\label{k19}\\
p_{1}  &  =-\frac{Mx}{2}\left(  1-\frac{p_{T}^{2}}{M^{2}x^{2}}\right)  ,\quad
p_{0}=\frac{Mx}{2}\left(  1+\frac{p_{T}^{2}}{M^{2}x^{2}}\right)  .\nonumber
\end{align}
These variables were used in our recent papers on TMDs
\cite{Zavada:2009sk,Efremov:2010mt}. Simply the value of invariant variable
$x$ does not depend on the reference frame, but its interpretation e.g. in the
rest frame differs from that in the infinite momentum frame.

\textit{ii)} The relations (\ref{k15}), (\ref{k15a}), which follow from RS,
can be confronted with the experimental data on $\left\langle p_{T}%
\right\rangle $ or $\left\langle \left\vert \mathbf{p}\right\vert
\right\rangle $. We have discussed the available data in \cite{Zavada:2009sk,
Efremov:2010mt} and apparently relation (\ref{k15a}) is compatible with the
set of lower values $\left\langle p_{T}\right\rangle $ corresponding to the
'leptonic data'. On the other hand the second set giving substantially greater
$\left\langle p_{T}\right\rangle $ and denoted as the 'hadronic data', seems
to contradict this relation. Actually the conflict with the relation
(\ref{k15a}) would mean either the conflict with some of the assumptions
\textit{a)--c) }or simply it can be due to absence of the higher order QCD
corrections. However, for possible comparison with the perturbative QCD
approach {\cite{arXiv:1101.5057}} let us remark the following. This approach
generate evolved TMDs using as input the existing phenomenological
parametrizations extracted from the experimental data. For example, one of the
inputs is the scale-independent Gaussian fit {\cite{arXiv:1003.2190}}%
\begin{equation}
F_{f/P}(x,p_{T})=f_{f/P}(x)\frac{\exp\left[  -p_{T}^{2}/\left\langle p_{T}%
^{2}\right\rangle \right]  }{\pi\left\langle p_{T}^{2}\right\rangle },
\label{k20}%
\end{equation}
where $\left\langle p_{T}^{2}\right\rangle =\left(  0.38\pm0.06\right)
\mathrm{GeV}^{2}$. Obviously our concept RS defined above is hardly compatible
with this distribution. In fact in the rest frame this distribution gives much
greater $\left\langle p_{T}^{2}\right\rangle $ than the corresponding
longitudinal $\ $term $\left\langle p_{1}^{2}\right\rangle $. However RS
requires $\left\langle p_{T}^{2}\right\rangle =2\left\langle p_{1}%
^{2}\right\rangle $ only. Let us remark that this imbalance is of the same
order as a difference between the two data sets mentioned above.

\textit{iii)} The \ first relation (\ref{k15}) apparently contradicts an
intuitive, Lorentz invariant condition%
\begin{equation}
\left\vert \mathbf{p}\right\vert ^{2}>p_{0}^{2} \label{k20a}%
\end{equation}
corresponding to bound, space-like quarks. Such conflict does not take place
for the leading order with quarks on-mass-shell. However, for off-shell quarks
the condition (\ref{k20a}) is incompatible with the assumptions
\textit{a)--c), }which imply (\ref{k15}). For example it means that any
approach, which is based primarily on the assumption\textit{ }(\ref{k20a}) in
some starting reference frame (including the infinite momentum frame) cannot
simultaneously satisfy the conditions \textit{a)--c)}.

\textit{iv)} In \cite{Zavada:2009sk} we explained why the RS, if applied on
the level of QPM, follows from the covariant description. In fact it means the
assumptions \textit{a)--c) }are common for our QPM and for the approaches like
\cite{D'Alesio:2009kv, Jackson:1989ph} where only Lorentz invariance is
explicitly required. The predictions of all these models are compatible with
the bound (\ref{k15a}). This is just a consequence of the fact that general
conditions \textit{a)--c) }are satisfied in these approaches. Another
theoretical reasons for RS have been suggested in \cite{Zavada:2001bq}. Let us
remark that the rotational-symmetry properties of the nucleon state in its
rest frame play an important role also in the recent studies
\cite{arXiv:1111.6069}.

\begin{figure}[ptb]
\includegraphics[width=6cm]{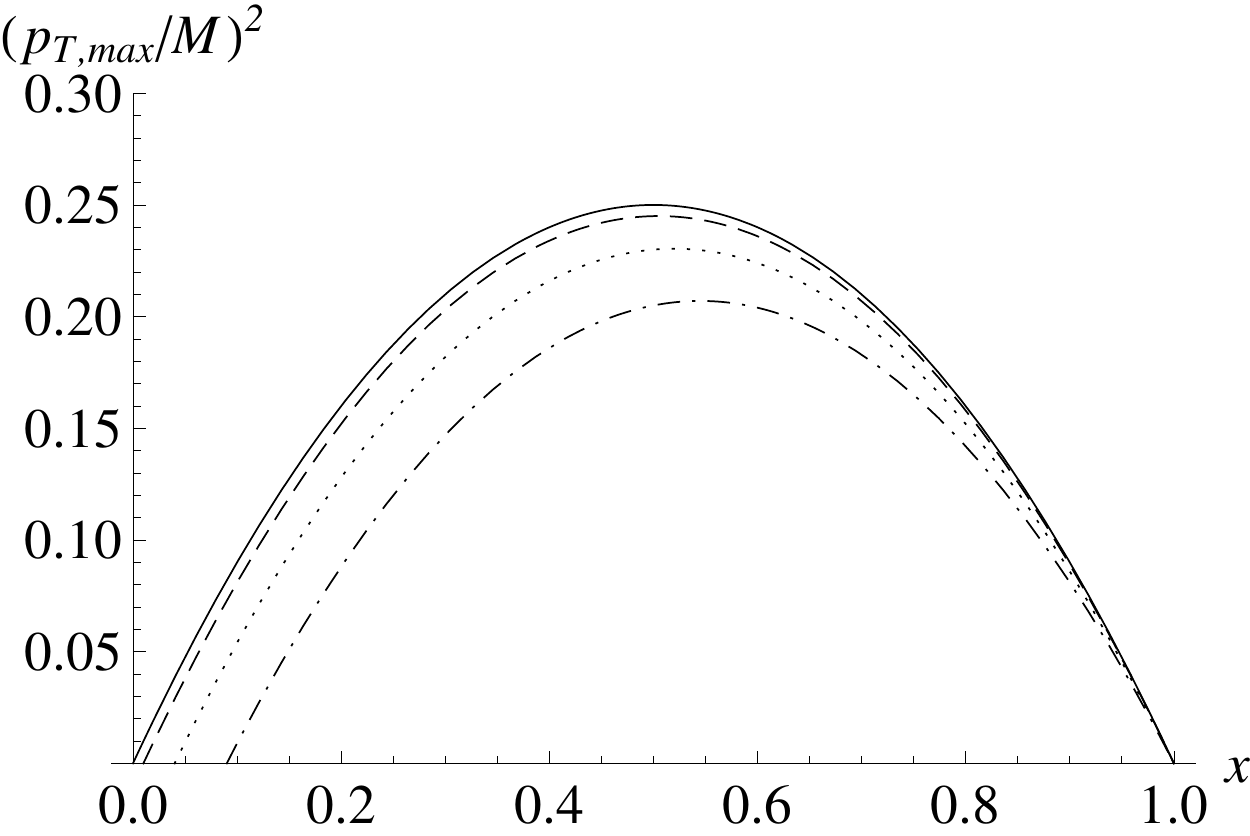}\caption{Upper limit of the quark
transversal momentum as a function of $x$ for $\mu=0$ (solid line), $0.1$
(dashed line), $0.2$ (dotted line) and $0.3$ (dash-dotted line). }%
\label{fe2}%
\end{figure}

\textit{v)} The relation (\ref{k17}) is obtained for the quarks on-mass-shell.
In a more general case, where only inequalities (\ref{k15}), (\ref{k15a})
hold, this relation is replaced by%
\begin{equation}
p_{T}^{2}\leq M^{2}\left(  x-\frac{\mu^{2}}{M^{2}}\right)  \left(  1-x\right)
;\qquad\mu^{2}\equiv p_{0}^{2}-\mathbf{p}^{2}, \label{k21}%
\end{equation}
where the term $\mu^{2}$ is not a parameter corresponding to the fixed mass,
but it is only a number varying in the limits defined by (\ref{k15}). The last
relation implies for any $\mu^{2}$:%
\begin{equation}
p_{T}^{2}\leq M^{2}x\left(  1-x\right)  , \label{k22}%
\end{equation}
which is equivalent to the on-mass-shell relation (\ref{k17}) for $m=0$. This
general upper limit for $p_{T}^{2}$\ depending on $x$ is displayed in Fig.
\ref{fe2}. Let us remark that results on $\left\langle p_{T}^{2}%
(x)\right\rangle $ obtained in \cite{Jackson:1989ph,D'Alesio:2009kv} are
compatible also with the bound (\ref{k22}). An equivalent form of this
inequality was probably for the first time presented in \cite{Sheiman:1979ku}.

\section{Summary}

In the present report we studied the kinematic constraints due to the
rotational symmetry of the quark momenta distribution inside the nucleon in
the leading order approach. In particular, we have shown that the light-cone
formalism combined with the assumption on the rotational symmetry in the
nucleon rest frame imply $p_{T}\leq M/2$. Only part of existing experimental
data on $\left\langle p_{T}\right\rangle $ satisfies this bound, but the
another part does not. In general, the existing methods for reconstruction of
$\left\langle p_{T}\right\rangle $\ from the DIS data are rather
model-dependent. These are the reasons why more studies are needed to clarify
these inconsistencies, since the phenomenological distributions in the
$x-p_{T}$ plane at present serve also as an input for a more fundamental
calculation of the QCD evolution and other effects.\smallskip

\begin{acknowledgments}
This work was supported by the project AV0Z10100502 of the Academy of Sciences
of the Czech Republic. I am grateful to Anatoli Efremov, Peter Schweitzer and
Oleg Teryaev for many useful discussions and valuable comments.
\end{acknowledgments}


\begin{thebibliography}{99}                                                                                               %


\bibitem {Efremov:2010mt}A.~V.~Efremov, P.~Schweitzer, O.~V.~Teryaev and
P.~Zavada,
Phys.\ Rev.\ D \textbf{83}, 054025 (2011).


\bibitem {Zavada:2009sk}P.~Zavada, Phys.\ Rev.\ D \textbf{83}, 014022 (2011)
[arXiv:0908.2316 [hep-ph]].




\bibitem {Efremov:2010cy}A.~V.~Efremov, P.~Schweitzer, O.~V.~Teryaev and
P.~Zavada,
PoS \textbf{DIS2010}, 253 (2010) [arXiv:1008.3827 [hep-ph]].


\bibitem {Efremov:2009ze}A.~V.~Efremov, P.~Schweitzer, O.~V.~Teryaev and
P.~Zavada,
Phys.\ Rev.\ D \textbf{80}, 014021 (2009).


\bibitem {Zavada:2007ww}P.~Zavada,
Eur.\ Phys.\ J.\ C \textbf{52}, 121 (2007).


\bibitem {Efremov:2004tz}A.~V.~Efremov, O.~V.~Teryaev and P.~Zavada,
Phys.\ Rev.\ D \textbf{70}, 054018 (2004).


\bibitem {Zavada:2002uz}P.~Zavada,
Phys.\ Rev.\ D \textbf{67}, 014019 (2003).


\bibitem {Zavada:2001bq}P.~Zavada,
Phys.\ Rev.\ D \textbf{65}, 054040 (2002).


\bibitem {Zavada:1996kp}P.~Zavada,
Phys.\ Rev.\ D \textbf{55}, 4290 (1997).




\bibitem {arXiv:1106.5607}P.~Zavada,
arXiv:1106.5607 [hep-ph] v1.




\bibitem {arXiv:1101.5057}S.~M.~Aybat and T.~C.~Rogers,
Phys.\ Rev.\ D\ \textbf{83}, 114042 (2011) [arXiv:1101.5057 [hep-ph]].




\bibitem {Aybat:2011ta} S.~M.~Aybat, A.~Prokudin and T.~C.~Rogers,
arXiv:1112.4423 [hep-ph].




\bibitem {arXiv:1003.2190}P.~Schweitzer, T.~Teckentrup and A.~Metz,
Phys.\ Rev.\ D\ \textbf{81}, 094019 (2010) [arXiv:1003.2190 [hep-ph]].




\bibitem {D'Alesio:2009kv}U.~D'Alesio, E.~Leader and F.~Murgia,
Phys.\ Rev.\ D \textbf{81}, 036010 (2010) [arXiv:0909.5650 [hep-ph]].


\bibitem {Jackson:1989ph}J.~D.~Jackson, G.~G.~Ross and R.~G.~Roberts,
Phys.\ Lett.\ B \textbf{226}, 159 (1989).




\bibitem {Ellis:1982cd}R.~K.~Ellis, W.~Furmanski and R.~Petronzio,
Nucl.\ Phys.\ B \textbf{212} (1983) 29;
Nucl.\ Phys.\ B \textbf{207}, 1 (1982).




\bibitem {arXiv:1111.6069}C.~Lorce' and B.~Pasquini,
arXiv:1111.6069 [hep-ph],
arXiv:1109.5864 [hep-ph],
arXiv:1107.3809 [hep-ph].



\bibitem {Sheiman:1979ku}J.~Sheiman,
Nucl.\ Phys.\ \textbf{B171}, 445 (1980).
\end{thebibliography}
\end{document}